\documentclass[prd,aps,10pt,twocolumn,showpacs,preprintnumbers,nofootinbib,superscriptaddress]{revtex4-1}
\usepackage{amsmath,amssymb,amsfonts,amsthm}
\usepackage{amsbsy} 
\usepackage{epsfig}
\usepackage{latexsym}

\usepackage{color}
\usepackage[breaklinks]{hyperref}

\def\com{\color{magenta}}
\def\cob{\color{blue}}
\newcommand{\oarX}[1]{\href{http://arxiv.org/abs/#1}{{\ttfamily\com #1}}}
\newcommand{\arX}[1]{\href{http://arxiv.org/abs/#1}{{\ttfamily\com arXiv:#1}}}
\newcommand{\doin}[2]{\href{http://dx.doi.org/#1}{\cob #2}}
\newcommand{\Eq}[1]{(\ref{#1})}

\def\barr{\begin{array}}
\def\earr{\end{array}}
\def\half{\frac{1}{2}}
\def\ben{\begin{equation}}
\def\een{\end{equation}}
\def\bs{\begin{subequations}}
\def\es{\end{subequations}}
\def\bena{\begin{eqnarray}}
\def\eena{\end{eqnarray}}

\def\bR{\mathbb{R}}

\def\bZ{\mathbb{Z}}
\def\cV{{\cal V}}
\def\im{i}
\def\cc{{\rm c.c.}}
\def\cK{{\cal K}}

\begin{document}

\title{Group field cosmology: a cosmological field theory of quantum geometry}
\author{Gianluca Calcagni}
\email{calcagni@aei.mpg.de}
\affiliation{Max Planck Institute for Gravitational Physics (Albert Einstein Institute), Am M\"uhlenberg 1, D-14476 Golm, Germany, EU}
\author{Steffen Gielen}
\email{sgielen@perimeterinstitute.ca}
\affiliation{Max Planck Institute for Gravitational Physics (Albert Einstein Institute), Am M\"uhlenberg 1, D-14476 Golm, Germany, EU}
\affiliation{Perimeter Institute for Theoretical Physics, 31 Caroline St. N., Waterloo, Ontario N2L 2Y5, Canada} 
\author{Daniele Oriti}
\email{doriti@aei.mpg.de}
\affiliation{Max Planck Institute for Gravitational Physics (Albert Einstein Institute), Am M\"uhlenberg 1, D-14476 Golm, Germany, EU}

\begin{abstract}
Following the idea of a field quantization of gravity as realized in group field theory, we construct a minisuperspace model where the wavefunction of canonical quantum cosmology (either Wheeler--DeWitt or loop quantum cosmology) is promoted to a field, the coordinates are minisuperspace variables, the kinetic operator is the Hamiltonian constraint operator, and the action features a nonlinear and possibly nonlocal interaction term. We discuss free-field classical solutions, the quantum propagator, and a mean-field approximation linearizing the equation of motion and augmenting the Hamiltonian constraint by an effective term mixing gravitational and matter variables. Depending on the choice of interaction, this can reproduce, for example, a cosmological constant, a scalar-field potential, or a curvature contribution.
\end{abstract}

\date{January 19, 2012}

\preprint{\doin{10.1088/0264-9381/29/10/105005}{Class.\ Quantum Grav.\ {\bf 29} (2012) 105005}\hspace{10cm}  \arX{1201.4151}}
\preprint{AEI-2011-099, pi-qg-255}

\maketitle


\section{Introduction and motivation}

Despite much recent progress \cite{libro}, background-independent approaches to quantum gravity face several open challenges. These concern: (i) the definition of the quantum dynamics of the fundamental degrees of freedom of spacetime that they identify, and the full control over it; (ii) the recovery of an effective description in terms of a smooth spacetime and geometry, once the dynamics is somehow defined, and in particular when the fundamental degrees of freedom are not continuous geometric data; (iii) the contact with the effective dynamics of general relativity and quantum field theory, and with phenomenology. 
 
An example is given by loop quantum gravity (LQG), a background-independent framework aiming to quantize the gravitational degrees of freedom in a nonperturbative way \cite{rovelli,thi01}. To this purpose, a canonical quantization scheme is employed where the constraints are written in terms of the densitized triad and of the Ashtekar--Barbero connection. The end result at the kinematical level is a Hilbert space of (spin network) states associated with graphs embedded in the spatial manifold and labeled by algebraic data (Lorentz group elements  or corresponding representation labels). As in any canonical scheme, while geometry is fully dynamical, the topology of the universe is fixed by construction, at least at the beginning. In general, however, one may ask whether it is possible to build a quantum theory inclusive of topology change or, in other words, if one can envisage an interacting multiverse scenario obeying a set of quantum rules. Since the degrees of freedom of a single universe are already fields, eventually to be quantized, such a scenario is sometimes said to be one of ``third quantization.'' This can be achieved, at least at a formal level, by defining a field theory over the space of geometries, for given spatial topology \cite{giddstrom,3rdquant}.\footnote{String field theory is an example of a ``third-quantized'' model. While the free Polyakov string is a collection of particle fields, a string field is a collection of strings interacting via certain vertices. One of the advantages to consider a field of strings is in the possibility to describe highly nonperturbative phenomena where the initial and final geometry and topology are different, such as brane decays into vacuum or into other branes (e.g., \cite{FK,cuta5,cuta7} and references therein).} Other (albeit inconclusive) arguments from a canonical quantum gravity perspective in favor of going to a ``third-quantization'' setting were also offered in \cite{kuchar,isham}.\footnote{To avoid confusion, from now on we employ the adjectives ``field'' or ``second'' instead of ``third'' to indicate this type of quantization.}

Beside the issue of topology change, the main difficulty faced by the LQG approach is the complete definition of the quantum dynamics and the proof that the resulting theory leads back to Einstein's gravity in an appropriate limit. 
A tentative but complete definition of the quantum dynamics of spin network states is obtained, via spin foam models \cite{Ori01,Per03} (a covariant definition of LQG dynamics), by embedding LQG states into the larger framework of group field theories (GFT) \cite{ori1,ori2,ori3}, in turn strictly related to tensor models \cite{tensor}. These are quantum field theories on group manifolds whose states are indeed spin networks and whose Feynman amplitudes are spin-foam models. This embedding has several advantages, from the LQG point of view. First of all, as said, it provides a complete definition of the quantum dynamics. Second, it defines such dynamics as the superposition of interaction processes (creation and annihilation) of spin-network vertices, forming complexes of arbitrary topology, such that topology is naturally made dynamical; it provides, in other words, a sort of local field-quantization scheme \cite{ori1,ori2}. Third, the field-theory framework offers powerful mathematical and conceptual tools for tackling the issue of the continuum limit and of the extraction of effective dynamics for better contact with phenomenology. In doing so, however, one has to abandon the familiar framework of canonical quantization of a classical (and local) field theory of gravitation, and is forced to face new types of conceptual and mathematical difficulties.

This program is just as ambitious as the original LQG one, if not more, and is difficult to realize in a complete and rigorous way, despite many recent advances. Toy models inspired by the full theory then become very important. In fact, they fulfill three main purposes: (i) they offer a simplified testing ground for ideas and techniques developed in the full theory; (ii) as such, they also have an important pedagogical value; (iii) they may represent, in principle, an effective, approximate framework to which the full quantum dynamics may reduce, in some limit, and thus they may be directly applicable to phenomenological studies. Obviously, due to their  simplicity, one should be cautious in interpreting the result obtained in the context of such toy models as truly physical, and their validity can be assessed only once the relation between toy model and full theory has been understood. 

An important type of simplified scenario has been developed in the context of LQG, in a symmetry-reduced setting of interest for cosmology. In fact, in order to understand certain features of loop quantum gravity, one often resorts to a minisuperspace model, loop quantum cosmology (LQC), where degrees of freedom are drastically reduced \cite{AsS,BCM}. In a pure Friedmann--Robertson--Walker (FRW) universe filled with a massless free scalar field, the classical and quantum dynamics of the universe as a whole can be described by the same formalism used for a free particle. In particular, the path integral is well defined \cite{ach1,ach2} and two-point correlation functions admit the usual classification \cite{2point}. By now, a wealth of interesting results have been obtained in this context \cite{AsS,BCM}. Given this analogy with the free particle, it is all the more natural to ask oneself if one can construct a sensible ``interaction'' among FRW universes and, once this is done, to change the interpretation of the two-point function from particle transition amplitude to field propagator as in the usual field quantization. One would then write down a field theory on minisuperspace, to obtain a field-quantized LQC framework. Another way to see the same field theory would then be as a toy model for group field theory, in which many of the difficult features of the latter are absent due to the global nature of the formulation and to the simplification provided by symmetry reduction, but where some ideas and techniques can still be applied. As with any toy model, one would then use it as a pedagogical testing ground and keep it available as a possible effective description of the full theory. 

We propose such a field theory for (loop) quantum cosmology in this paper, with the above motivations. The presentation is organized as follows. We review some basic features of LQC in Sec.\ \ref{loqc}, but the Wheeler--DeWitt case is also easily recovered. In Sec.\ \ref{gufc} the field theory is defined by promoting the quantum Hamiltonian constraint to the kinetic operator of a (real) scalar field $\Psi$ on minisuperspace. We analyze the relation between different kinetic operators and the gauge choice, with particular focus on exactly solvable free theories. We discuss the various possible choices for the interaction term. Following this general definition, we move on to analyze some consequences of the formalism. We analyze the free propagator of the theory first, corresponding to the evolution of a single universe (Sec.\ \ref{fft}). We show how the embedding into a field-theory setting has immediate interesting consequences also for the single-universe dynamics. Then, we consider how the presence of interactions affects this single-universe evolution. Approximating the interaction as a mean-field term, we find an effective equation linear in the field $\Psi$, correcting the Hamiltonian quantum constraint equation by an extra term (Sec.\ \ref{mfa}). The latter mixes, in general, gravitational and matter degrees of freedom, and its exact form depends on the chosen initial interaction as well as on the mean-field configuration considered. We conclude with a discussion of other possible applications of this formalism.


\section{Brief overview of loop quantum cosmology} \label{loqc}

\subsection{Classical theory}

Our starting point is the description within loop quantum cosmology of the spatially flat, homogeneous and isotropic universe with a massless scalar field as matter, which we summarize briefly in this section. In the canonical analysis of dimensionally-reduced general relativity, one restricts integrations to a fixed fiducial three-dimensional cell of comoving volume $\cV_0<\infty$, with a flat metric ${}^0 q_{ab}$ which may be taken to be $\delta_{ab}$ in Cartesian coordinates. The four-dimensional metric is then
\ben
ds^2 = -N^2(t)\, dt^2 + a^2(t)\,{}^0 q_{ab}\,dx^a\,dx^b\,,
\label{frwmetric}
\een
where $a(t)$ is the scale factor, spatial indices are labeled by Latin indices $a,b,\dots=1,2,3$, and there is a freedom in the choice of the lapse function $N(t)$. Indices $i,j,\dots=1,2,3$ will denote directions in the tangent space.

With a choice of frame $\{{}^0 e^a_i\}$ and dual $\{{}^0 e_a^i\}$, orthonormal with respect to ${}^0 q_{ab}$, the physical triad $e^i=\varepsilon\,a\,{}^0 e_a^i\,dx^a$ ($\varepsilon=\pm 1$) and $e^0=N\,dt$ are orthonormal with respect to (\ref{frwmetric}), and the Levi-Civita connection is
\ben
{\omega^i}_0 = \varepsilon\,\frac{\dot{a}}{N}\,{}^0 e^i\,,\qquad {\omega^i}_j=0\,,
\label{levi}
\een
where a dot denotes time derivative. From this one computes the variables used in loop quantum gravity, the Ashtekar--Barbero $\mathfrak{su}(2)$ connection $A_a^i$ and the densitized triad $E^a_i$, via
\bs\label{lqgvaria}\bena
A_a^i &=& \gamma\,\left({\omega^i}_0\right)_{a}\,,\\
E^a_i &=&(\det e)e^a_i = a^2\sqrt{\det {}^0 q}\,{}^0 e_i^a\,,
\eena\es
where $\gamma$ is the Barbero--Immirzi parameter.

A shortcut to the standard canonical analysis is to substitute the FRW metric and the Ricci scalar stemming from (\ref{frwmetric}),
\ben
R=6\left(\frac{\ddot{a}}{a\,N^2}-\frac{\dot{a}\dot{N}}{aN^3}+\frac{\dot{a}^2}{a^2 N^2}\right)\,,
\een
into the Einstein--Hilbert and matter action, which then depend on $\phi$, $a$, and $N$. The conjugate momenta are $p_a=- 3 \cV_0\, a\dot{a}/(4\pi G \, N)$ and $p_\phi=\cV_0\,a^3\dot{\phi}/N$, and the conservation in time of the primary constraint $p_N\approx 0$ (the symbol $\approx$ denotes weak equality) leads to the Friedmann equation
\ben
\cK:=\frac{2\pi G}{3}\frac{p_a^2}{a}-\frac{p_\phi^2}{2a^3}\approx 0\,,
\label{friedm}
\een
which should be imposed as a constraint on quantum states in quantum cosmology.\footnote{Throughout the paper we use the symbol $\cK$ for the Hamiltonian constraint because it will eventually be regarded as a kinetic operator. Although this differs from the more standard choice of symbol $H$ or ${\cal H}$, it has the further advantage of avoiding confusion with the Hubble parameter.}

\subsection{Kinematics}

Focusing on the gravitational sector for now, the crucial difference from traditional minisuperspace (Wheeler--DeWitt) approaches to quantum cosmology in LQC is that one follows the kinematics of full loop quantum gravity, where not the connection but only its holonomies are defined as operators \cite{rovelli}. It is convenient to introduce new conjugate variables $c$ and $p$, where
\ben
c=\varepsilon\,\cV_0^{1/3}\,\frac{\gamma\,\dot{a}}{N}=-\varepsilon\,\frac{4\pi G\,\gamma}{3\cV_0^{2/3}}\,\frac{p_a}{a}\,,\quad p=\varepsilon\,a^2\cV_0^{2/3}\,,
\een
so that $A_a^i$ only depends on $c$, and powers of $\cV_0$ have been introduced to make $c$ and $p$ invariant under the residual symmetry $a\rightarrow\lambda\,a$, ${}^0 q_{ab}\rightarrow\,{}^0 q_{ab}/\lambda^2$ in Eq.~(\ref{frwmetric}). Instead of $\hat{c}$ and $\hat{p}$ one now defines $\hat{p}$ and $\widehat{\exp({\rm i}\mu c)}$ as operators, where $\mu$ can be a real parameter or a function of $p$ chosen by means of a suitable procedure.

The kinematical Hilbert space $\mathcal{H}_{{\rm kin}}^{{\rm g}}$ is taken to be the space of square-integrable functions on the Bohr compactification of the real line. One can work in a basis where $\hat{p}$ is diagonal, with orthonormality relation $\langle p | p' \rangle = \delta_{p,p'}$, so that one is dealing with a nonseparable Hilbert space. In this representation, if $\mu$ is taken to be a nontrivial function of $p$ the action of the holonomy operator $\widehat{\exp({\rm i}\mu c)}$ takes a rather complicated form, and it is convenient to choose a different representation. In the improved dynamics scheme \cite{improv}, where $\mu(p)\sim|p|^{-1/2}$, this is a basis $\{|\nu\rangle\}$ of eigenstates of the volume operator $\hat{\cV}$ measuring the kinematical volume of the fiducial cell, $\cV=|p|^{3/2}$,
\ben
\hat{\cV}|\nu\rangle = 2\pi\gamma G\, |\nu|\, |\nu\rangle\,,
\een
where $\nu=\varepsilon a^3 \cV_0/(2\pi \gamma G)$ has dimensions of length. The states $\{|\nu\rangle\}$ can be normalized to
\ben
\langle \nu | \nu' \rangle = \delta_{\nu,\nu'}\,.
\een
The basic operators are now $\hat\nu$, which acts by multiplication, and $\widehat{\exp(\im\lambda b)}$, where $b=\varepsilon\,(2\pi\gamma G\, p_a)/(3\cV_0\,a^2)$ is conjugate to $\nu$ [and is proportional to the Hubble parameter $H=\dot a/(N a)$] and $\lambda={\rm const}$, which acts as a shift in $\nu$. These satisfy the standard Heisenberg algebra. For the matter sector, one chooses the usual Schr\"odinger quantization with a natural representation of the Hilbert space $\mathcal{H}_{{\rm kin}}^{\phi}$, the space of square-integrable functions on $\bR$, on which $\hat{\phi}$ acts by multiplication and $\hat{p_{\phi}}$ by derivation, and with an orthonormal basis given by 
\ben\label{phino}
\langle \phi | \phi' \rangle = \delta(\phi-\phi')\,.
\een
The Hilbert space of the coupled system is then just the tensor product $\mathcal{H}_{{\rm kin}}^{{\rm g}} \otimes \mathcal{H}_{{\rm kin}}^{\phi}$. As in traditional approaches to quantum cosmology, the variable $N$ is removed from the configuration space because the primary constraint $p_N\approx 0$ would mean that wavefunctions are independent from $N$. We note that the full constraint would be a multiple of $N$, so that in situations where the resulting quantum constraint depends on the choice of lapse function (as below) the choice $N=1$ seems more natural when considering that $N$ is also originally in the configuration space. In fact, in cosmology the lapse can be regarded as a function of the scale factor, $N=N(a)$, which is an independent variable. A choice of the form $N=N(E)$ in the full theory is somewhat less justified before solving the constraints.

\subsection{Dynamics}

The quantum analogue of the Friedmann equation (\ref{friedm}) is obtained by starting with the Hamiltonian constraint of full general relativity in terms of the variables (\ref{lqgvaria}) and expressing the curvature of $A_a^i$ through the holonomy around a loop, taking account of the {\it area gap} ---the result in LQG that the area of such a loop cannot assume arbitrarily small nonzero values. The Hamiltonian constraint is
\ben
\hat{\cK}\,\psi(\nu,\phi) := -B(\nu)\left(\Theta+\partial_{\phi}^2\right)\psi(\nu,\phi)= 0\,,
\label{wdw}
\een
where $\psi$ is a wavefunction on configuration space and $\Theta$ is a difference operator only acting on $\mathcal{H}_{{\rm kin}}^{{\rm g}}$ and of the form
\bena
-B(\nu)\Theta\psi(\nu,\phi) &:= &A(\nu)\psi(\nu+\nu_0,\phi)+C(\nu)\psi(\nu,\phi)\nonumber
\\& & +D(\nu)\psi(\nu-\nu_0,\phi)\,,
\label{thet}
\eena
where $A,B,C$, and $D$ are functions which depend on the details of the quantization scheme ({\it inter alia}, on the choice of lapse function) and $\nu_0$ is an elementary length unit, usually defined by the square root of the area gap (the Planck length up to a numerical factor). The physical states are the solutions of Eq.~(\ref{wdw}). Due to the structure of Eq.~(\ref{thet}), in LQC one has an interval's worth of superselection sectors in $\mathcal{H}_{{\rm kin}}^{{\rm g}}$: $\Theta$ preserves all subspaces spanned by $\{|\nu_I+n\nu_0\rangle\,|\,n\in\bZ\}$ for some $\nu_I$. We may restrict ourselves to one of these subspaces, i.e., assume that wavefunctions only have support on a discrete lattice which we take to be $\nu_0\bZ$ [for a generic gauge choice, there may be issues with the definition of (\ref{wdw}) at the most interesting point $\nu=0$]. This restriction picks out a separable subspace to which we will limit our analysis. 


\section{Defining the field theory}\label{gufc}

We now define our field theory on (mini)superspace. The Hamiltonian constraint (\ref{wdw}) of the first-quantized theory is the natural starting point for the free action of the field theory. We define this action to be
\ben
S_{{\rm f}}[\Psi]=\half\sum_{\nu}\int d\phi\;\Psi(\nu,\phi)\hat{\cK}\Psi(\nu,\phi)\,,
\label{akshn}
\een
where in the simplest setting we take $\Psi$ to be a real scalar field. If $\hat \cK$ is as in Eq.~\Eq{thet}, we must assume that the combination $B(\nu)\Theta$ is symmetric in $\nu$, i.e., that
\ben\label{da}
D(\nu)=A(\nu-\nu_0)
\een
in Eq.~(\ref{thet}), in order to reproduce the equation of motion (\ref{wdw}). Put differently, for any constraint (\ref{thet}) the action (\ref{akshn}) projects out its self-adjoint part with respect to the measure given by the kinematical inner product of LQC. Taking this measure as given, possible manipulations of the constraint $\hat{\cK}$ are restricted by this requirement. Notice, however, that Eq.~\Eq{da} does hold in LQC for the usual choices of gauge, so we do not need to impose it as an additional requirement. To give an example, for the preferred lapse choice $N=1$ and in improved dynamics, the functions $A,B,C$ take the form \cite{improv}
\bena
&&A(\nu)=\frac{1}{12\gamma\sqrt{2\sqrt{3}}}\left|\nu+\frac{\nu_0}{2}\right|\left| \left|\nu+\frac{\nu_0}{4}\right|-\left|\nu+\frac{3\nu_0}{4}\right| \right|\,,\nonumber
\\&&B(\nu)=\frac{3\sqrt{2}}{8\sqrt{\sqrt{3}}\pi\gamma G}|\nu| \left| \left|\nu+\frac{\nu_0}{4}\right|^{\frac{1}{3}}-\left|\nu-\frac{\nu_0}{4}\right|^{\frac{1}{3}} \right|^3\,,\nonumber
\\&&C(\nu)=-A(\nu)-A(\nu-\nu_0)\,,
\label{complicated}
\eena
where now $\nu_0=4\lambda:=\sqrt{32\sqrt{3}\,\pi\gamma\,G}$, whereas for the solvable ``sLQC'' model in \cite{acs} [which uses $N=a^3$ and the symmetry (\ref{symmetry})],
\bena
&&A(\nu)=\frac{\sqrt{3}}{8\gamma}\left(\nu+\frac{\nu_0}{2}\right)\,,\quad B(\nu)=\frac{1}{\nu}\,,\nonumber
\\&&C(\nu)=-A(\nu)-A(\nu-\nu_0)=-\frac{\sqrt{3}}{4\gamma}\nu\,.
\label{slqcfunc}
\eena
Equation (\ref{slqcfunc}) can be shown to agree with the previous expressions (\ref{complicated}) in the ``semiclassical'' limit $\nu \gg \nu_0$.

In LQC, one normally assumes symmetry of the wavefunction $\psi$ under orientation reversal,
\ben
\psi(\nu,\phi)=\psi(-\nu,\phi)\,,
\label{symmetry}
\een
since the kinematical Hilbert space can be split into symmetric and antisymmetric subspaces which are superselected (in other words, the physics should not depend on the frame orientation). From the field theory perspective, such a requirement is less natural, in particular if interactions are taken into account; we will allow for general field configurations without assuming Eq.~(\ref{symmetry}).

By definition of second quantization, and by construction in our case, the classical solutions of the free field theory will correspond to the quantum solutions of the first-quantized model. 

\

We now complete the definition of the field theory on minisuperspace with the addition of an interaction term for our field. The first-quantized theory, that is (loop) quantum cosmology, does not offer indications on how this interaction should be defined, so one has to proceed in a rather exploratory way guided only by general intuition (and by the results obtained following various choices). One could take an arbitrary functional, but we opt for an $n$th-order polynomial in $\Psi$ not necessarily local in the minisuperspace variables.  We will specialize to simpler, concrete choices in the following, in order to study some consequences of the formalism. 

We get the general form for the interacting theory
\bena
S_{{\rm i}}[\Psi]&=&\half\sum_{\nu}\int d\phi\;\Psi(\nu,\phi)\hat{\cK}\Psi(\nu,\phi)+\sum_{j=2}^n\frac{\lambda_j}{j!}\times\label{intact}
\\&&\sum_{\nu_1\ldots\nu_j}\int d\phi_1\ldots d\phi_j\; f_j(\nu_i,\phi_i) \prod_{k=1}^j\Psi(\nu_k,\phi_k)\,,\nonumber
\eena
where $f_j(\nu_i,\phi_i)$ are unspecified functions depending on $\{\nu_i,\phi_i\}_{i=1,\ldots, j}$. This gives the equation of motion
\bena
\hat{\cK}\Psi(\nu,\phi)+\sum_{j=2}^n \frac{\lambda_j}{j!}\sum_{\nu_1\ldots\nu_{j-1}}\int d\phi_1\ldots d\phi_{j-1}&&
\label{fieldeq}
\\ \times\prod_{l=1}^{j-1}\Psi(\nu_l,\phi_l)\sum_{k=1}^j \hat{f}_k(\nu_i,\phi_i;\nu,\phi) &=&0,\nonumber
\eena
where 
\ben
\hat{f}_k(\nu_i,\phi_i;\nu,\phi) := f(\mu_i,\varphi_i)\,,
\een
with $\{\mu_i\}=\{\nu_1,\ldots,\nu_{k-1},\nu,\nu_k,\ldots,\nu_j\}$ and $\{\varphi_i\}=\{\phi_1,\ldots,\phi_{k-1},\phi,\phi_k,\ldots,\phi_j\}$. 
In writing down the field theory action we have included possible nonlocal (in $\nu$) quadratic terms in the interaction part (specified by the interaction kernel $f_2$) rather than in the kinetic term, in order to emphasize the fact that $\hat{\cK}$ is usually chosen to be a local operator in the geometry in quantum cosmology. These could however be also included in the definition of $\hat{\cK}$.

\

The above is rather general. Explicit analyses require choosing specific interaction terms, that is, functions $f_j(\nu_i,\phi_i)$. Each of them could correspond to a choice of a physical quantity to be conserved through the ``interaction'' of the universes, and of a conjugate physical quantity in terms of which the fields interact, instead, ``locally.'' That means that the conjugate variable is identified across ``incoming'' and ``outgoing'' universes.  Different choices for the function $f(\nu_i,\phi_i)$, and hence for the quantities conserved in the interaction, can thus be motivated by different physical considerations. In particular, the choice will be influenced by the interpretation of such an interaction as true topology change, that is, splitting/merging of universes, or rather as a merging/splitting of homogeneous and isotropic patches within a single inhomogeneous and anisotropic universe. The detailed definition of the second scenario in physical terms is not easy, and we will confine our treatment to a brief discussion of it at the end. However, it is important to keep in mind that its consideration would sensibly affect the very definition of the field theory to analyze. Examples of interactions are:
\begin{itemize}
\item If, e.g., only $\lambda_3$ is nonvanishing and we implement locality in ($\nu,\phi$) by choosing $f(\nu_i,\phi_i)=\delta_{\nu_1,\nu_2}\delta_{\nu_1,\nu_3}\delta(\phi_1-\phi_2)\delta(\phi_1-\phi_3)$, we have the field equation
\ben
\hat{\cK}\Psi(\nu,\phi)+\frac{\lambda_3}{2}\Psi^2(\nu,\phi)=0\,.
\een
Such a potential, which implements locality both in the 3-volume (i.e., the scale factor) and in the scalar field $\phi$, implies the existence of conservation laws for the conjugate quantities $b$ and $p_\phi$. These conservation laws are nothing but the (modified) second Friedmann and Klein--Gordon equations, respectively. Since interactions allow for topology change, in this case for nonzero $\lambda_3$ there is a process where two ``universes'' merge into one, and the metric in both ingoing patches as well as in the outgoing patch is required to be the same (analogous to the interaction term proposed in \cite{giddstrom}). Because of the conservation law, there is a discontinuity in the Hubble parameter $b\propto H$.
\item A different conservation law, namely conservation of Hubble volume $b^{-3}$ or locality in the conjugate quantity $b^4\nu$, is suggested if one interprets the interaction as the topology-changing process just described with a discontinuity in the causal past of an observer passing the ``merging point'' (this is typically considered ``bad'' topology change; see the review \cite{dowker}). 
\item Another possibility would be to take the conserved quantities of the classical (Wheeler--DeWitt) Hamiltonian $p_\phi$ and $\nu b$ as also conserved in interactions, which would then be local in $\phi$ and $\ln(\nu/\nu_0)$. For the sLQC model detailed below, the second quantity would be modified to $(2\nu/\nu_0)\sin\left(b\nu_0/2\right)$ and we would require locality in the conjugate variable $\ln\{(\nu/\nu_0)[1+\cos(b\nu_0/2)]\}$.
\end{itemize}

In the Wheeler--DeWitt case, we can choose the type of interaction analogously, since the only difference is in the kinetic operator (choice of first-quantization scheme) and not in the choice of minisuperspace variables.

\

Before starting our analysis of the model, we introduce a reformulation of the same that is advantageous for practical manipulations.

Since $\hat{\cK}$ is not diagonal in $\nu$, it may be convenient to use a Fourier transform as in \cite{acs},
\bena
&&\Psi(b,\phi):=\sum_{\nu}e^{\im\nu b}\Psi(\nu,\phi),
\\&&\Psi(\nu,\phi)=\frac{\nu_0}{2\pi}\int\limits_0^{2\pi/\nu_0} db\,e^{-\im\nu b}\Psi(b,\phi)\,.
\eena
As $\Psi(\nu,\phi)$ is real, $\Psi(b,\phi)=\overline\Psi(2\pi/\nu_0-b,\phi)$, and if Eq.~(\ref{symmetry}) holds, then $\Psi(b,\phi)=\Psi(2\pi/\nu_0-b,\phi)$. The action (\ref{intact}) becomes
\bena
S_{{\rm i}}[\Psi]&=&\frac{\nu_0}{4\pi}\int d\phi\int\limits db\,\overline\Psi(b,\phi)\hat{\tilde{\cK}}\Psi(b,\phi)+\sum_{j=2}^n\frac{\lambda_j}{j!}
\\&&\times\left(\frac{\nu_0}{2\pi}\right)^j\int \prod_{l=1}^j d\phi_l\,db_l\,\bar{f}(b_i,\phi_i)\prod_{l=1}^j\Psi(b_l,\phi_l)\,,\nonumber
\eena
where $\bar{f}(b_i,\phi_i):=\sum_{\nu_1\ldots\nu_j}e^{-\im\sum_k \nu_k b_k}f(\nu_i,\phi_i)$ is the (complex conjugate of the) Fourier transform of the function $f(\nu_i,\phi_i)$ appearing in (\ref{intact}), and $\hat{\tilde{\cK}}$ is an appropriate differential operator. In the above conventions,
\bena
\hat{\tilde{\cK}}&=&e^{\im\nu_0 b}A\left(-\im\partial_b\right)+C\left(-\im\partial_b\right)+A\left(-\im\partial_b\right)e^{-\im\nu_0 b}\nonumber
\label{cofb}
\\&&+B\left(-\im\partial_b\right)\partial_{\phi}^2\,.
\eena
The inverse of this operator will give the propagator. 

In the sLQC case, Eq.~(\ref{slqcfunc}), in order to avoid nonlocal expressions such as $1/(-\im \partial_b)$ in the action we could multiply the expression for $\Theta$ by $\nu$. However, this would lead to a ``nonsymmetric'' form such that Eq.~\Eq{da} is not respected [obviously, $\nu D(\nu)=\nu A(\nu-\nu_0)\neq (\nu-\nu_0) A(\nu-\nu_0)$] and $\hat \cK$ is not self-adjoint. A symmetrized version of $\hat \cK$, which has not been previously considered in the literature, has 
\ben
A(\nu)=\frac{\sqrt{3}}{8\gamma}\left(\nu+\frac{\nu_0}{2}\right)^2,\; B(\nu)=1,\; C(\nu)=-\frac{\sqrt{3}}{4\gamma}\nu^2.
\label{symme}
\een
When the nonsymmetric form resulting from multiplication by $B^{-1}(\nu)$ is used in LQC, the kinematical inner product is modified accordingly to keep the constraint self-adjoint (see, e.g., \cite{selfad}). The action (\ref{akshn}), which involves both the constraint and the kinematical inner product of LQC, is left unchanged by such a redefinition.

\

As mentioned above, the LQC setting is chosen because of the initial motivation to obtain a toy model for a GFT construction, in turn a field theory formalism for LQG states. This is, however, not essential for our general purposes. One could define an analogous set of models for Wheeler--DeWitt quantum cosmology, which is actually simpler from a technical point of view. The usual ordering for the quantum Friedmann equation is \cite{hawpage}
\ben
\hat{\cK}:= \frac{4\pi G}{3}\left(a\,\partial_a\right)^2-p_\phi^2\,,
\label{wdwcon}
\een
and the scale factor $a$ is now a continuous variable. It is then convenient to define ${\cal N}:=\sqrt{3/(4\pi G)}\ln a$ so that $a$ is restricted to be nonnegative and the constraint is simply the Klein--Gordon operator $\partial_{{\cal N}}^2-\partial_\phi^2$. One ends up with a scalar field theory in 1+1 dimensions with standard kinetic operator and an unusual potential term.

\subsection{Fock space construction}

As any ordinary field theory, the above field theory on minisuperspace can be quantized and a Fock space of quantum states can be constructed. The construction of the Fock space rests on a choice of a complete basis on the space of fields. 

As customary, we can choose a basis of solutions of the free field theory, that is single-particle states in $\{|k\rangle\}$ whose elements correspond to classical (expanding or contracting) solutions of the modified Friedmann dynamics.\footnote{We warn the reader about a fine point in terminology. In the LQC and Wheeler--DeWitt literature, by ``classical solutions'' one often means the solutions of the unmodified Friedmann equations in classical general relativity. In contrast, by ``classical trajectories'' we presently mean solutions generated by a Hamiltonian that already includes the two effects of replacing the connection by its holonomies and of setting a minimal area for closed holonomies [such as Eq.~\Eq{moh}]. Both operations can be motivated by the first-quantization framework of LQC, but they are performed already at the classical level.} These solutions are labeled by $k$, i.e., 3-geometries (described by $\nu$) embeddable into a four-dimensional FRW universe by means of $k$ and a choice of lapse $N(t)$. In solvable LQC, the deparametrized solutions are of the form
\ben\label{lit}
b(\phi)=\frac{4}{\nu_0}\arctan\exp\left[\pm\sqrt{12 \pi G}(\phi-\phi_0)\right]\,,
\een
and $k$ is the value of the Dirac observable $p_\phi$. Equation \Eq{lit}, specifying the classical trajectories, will later reappear as the ``light cone'' $x(b)=\phi$ of the propagator of the theory, see Eq.~\Eq{prpx}. In the gauge $N=a^3$, consider the modified Hamiltonian
\ben\label{moh}
\cK=\frac{p_\phi^2(t)}{2}-6\pi
G\left\{\nu(t)\frac{2}{\nu_0}\sin\left[\frac{b(t)\nu_0}{2}\right]\right\}^2\,.
\een
Hamilton's equations give $\dot\phi=p_\phi$ and $\dot{\nu}=\partial \cK/\partial b$, solved by
\bena
b(t)&=&\frac{4}{\nu_0}\arctan\exp\left[\pm\sqrt{12 \pi G}k(t-t_0)\right]\,,
\\\nu(t)&=&\pm\frac{k\nu_0\cosh[\pm\sqrt{12\pi G}\,k(t-t_0)]}{\sqrt{48\pi G}}\,,\label{cosh}
\eena
so that the bounce is apparent already in each classical history (we have treated $\nu$ as a continuous parameter for the purpose of simplicity here). In fact, Eq.~\Eq{cosh} is consistent both with \Eq{lit} ($\phi=\pm k t$ are solutions of the constraint $p_\phi={\rm const}$) and with the expectation value of the volume operator \Eq{coshv}.

The construction of the Fock space proceeds as usual. One defines creation operators $a_k^{\dagger}$ and annihilation operators $a_k$ from the mode decomposition of generic fields into the above basis, and builds generic elements of the Hilbert space of states from their combined action on the Fock vacuum state $| 0\rangle$. 

This state, as in the complete GFT formalism \cite{ori1,ori2}, would correspond to a very degenerate ``no-space'' state, in which no geometric and no topological structure at all is present. Topological and geometric structures are created out of it by the action of the creation operators.  

A crucial ingredient in the definition of the Fock space is the choice of quantum statistics. In the following, we fix the statistics to be bosonic, $[a_k,a_{k'}^{\dagger}]\propto \delta_{k,k'}$. This seems physically natural if quantum states are associated with classical geometries and whole universes. 

In group field theory, where the Fock space would be constructed out of microscopic ``building blocks'' of space \cite{ori2}, the choice of statistics is less obvious and it is the focus of current research (see for example the discussion in \cite{GFTdiffeos}). We may also expect the situation to be subtler if the objects created and annihilated in our field theory are to be interpreted as local homogeneous and isotropic patches of a single universe. However, as anticipated, we do not discuss in detail this possibility in this work, and thus stick to the simplest choice of statistics.

\section{The free field theory}\label{fft}
We now begin the analysis of the field theory we defined. We limit most of our considerations to solvable sLQC, where explicit calculations can be performed. However, we try to maintain a certain level of generality in the presentation, to leave room for the study of other cases. We start with the analysis of the free field theory. We have seen that, already at this level, the field-theory setting implies certain restrictions on and modifications of the LQC Hamiltonian constraint operators, and thus of the single-universe dynamics. We focus on these first.

\subsection{Equations of motion and Hamiltonian constraint}
For sLQC, the free equation (\ref{wdw}) can be solved analytically. We reserve the symbol $\psi$ for field solutions of the free theory and the symbol $\Psi$ for the classical field solutions of the interacting model. Inserting Eq.~(\ref{slqcfunc}) into (\ref{cofb}), one obtains
\ben
\im\partial_b\hat{\tilde{\cK}}\psi=\left\{\partial_\phi^2-12\pi G\left[\partial_b\frac{2}{\nu_0}\sin\left(\frac{b\,\nu_0}{2}\right)\right]^2\right\}\psi\,.
\label{slqc}
\een
Setting $\psi=\partial_b\chi$ and choosing an (irrelevant) integration constant to be zero yields
\ben
\hat{\tilde{\cK}}\chi=-\im\left\{\partial_\phi^2-12\pi G\left[\frac{2}{\nu_0}\sin\left(\frac{b\,\nu_0}{2}\right)\partial_b\right]^2\right\}\chi=0\,,
\een
with general solution $\chi=\chi_+[\phi-x(b)]+\chi_-[\phi+x(b)]$, where 
\ben
x(b)=\frac{1}{\sqrt{12\pi G}}\ln\left[\tan\left(\frac{\nu_0 b}{4}\right)\right]\,,
\label{transf}
\een
so that $x'(b)=[(2\sqrt{12\pi G}/\nu_0)\sin(\nu_0 b/2)]^{-1}$. Note that $b$ parametrizes the circle $S^1$ from which the point $b=0$ (or $b=2\pi/\nu_0$) must be removed to make the coordinate transformation (\ref{transf}) well defined. As $b\rightarrow 0$, we see that $dx/db\rightarrow\infty$. We will encounter this singular behavior when giving the expression of the propagator of the theory in terms of $b$, see Eq.\ (\ref{prpx}) below.

We may expand the general solution in Fourier modes,
\ben
\chi(b,\phi)=\int\limits_{-\infty}^{{+\infty}}dk\,\left\{\tilde{A}_{\rm L}(k)\,e^{\im k[\phi-x(b)]}+\tilde{A}_{\rm R}(k)\,e^{\im k[\phi+x(b)]}\right\},\label{chi}
\een
and hence
\bena
\psi(\nu,\phi)&=&\frac{\nu_0}{2\pi}\int\limits_0^{2\pi/\nu_0} db\,e^{-\im\nu b}\int\limits_{-\infty}^{+\infty}dk\,x'(b)\nonumber
\\&&\times\left\{A_{\rm L}(k)\,e^{\im k[\phi-x(b)]}+A_{\rm R}(k)\,e^{\im k[\phi+x(b)]}\right\}\nonumber
\\&=&\frac{\nu_0}{2\pi}\int\limits_{-\infty}^{+\infty} dx\,e^{-\im\nu b(x)}\int\limits_{-\infty}^{+\infty}dk
\label{solution}
\\&&\times\left[A_{\rm L}(k)\,e^{\im k(\phi-x)}+A_{\rm R}(k)\,e^{\im k(\phi+x)}\right]\,,\nonumber
\eena
where $b(x)=(4/\nu_0)\arctan \exp(\sqrt{12 \pi G}x)$ and $A_{\rm L}:=-\im k\tilde{A}_{\rm L}$ and $A_{\rm R}:=\im k\tilde{A}_{\rm R}$ are chosen so that $\psi$ is real.

In LQC (first-quantized theory), the existence of a bounce can be proven analytically with the state $\chi$. Noting that the volume operator in $x$ representation acts as $\hat\nu\propto \partial_b\propto \cosh(\sqrt{12\pi G} x)\partial_x$, one can compute its expectation value on the state \Eq{chi}, with the scalar product
\ben
\langle\chi| \hat\nu|\chi\rangle=-i\int dx (\chi^*\partial_\phi|\hat\nu|\chi-|\hat\nu|\chi\partial_\phi\chi^*)\,.
\een
Since $\chi$ obeys a Klein--Gordon equation in $\phi$ and $x$ and its left and right sectors depend on the combinations $\phi\pm x$, it is easy to see that the cosh in the volume operator factorizes as a cosh in the scalar field, so that 
\ben\label{coshv}
\langle\chi| \hat\nu|\chi\rangle=\nu_*\cosh(\sqrt{12\pi G} \phi)\,,
\een
where the proportionality coefficient $\nu_*$ is the minimum volume at the bounce. Later on we shall derive this result again from the ``light-cone'' condition of the field-theory propagator.

\begin{figure}[ht]
\includegraphics[scale=0.67]{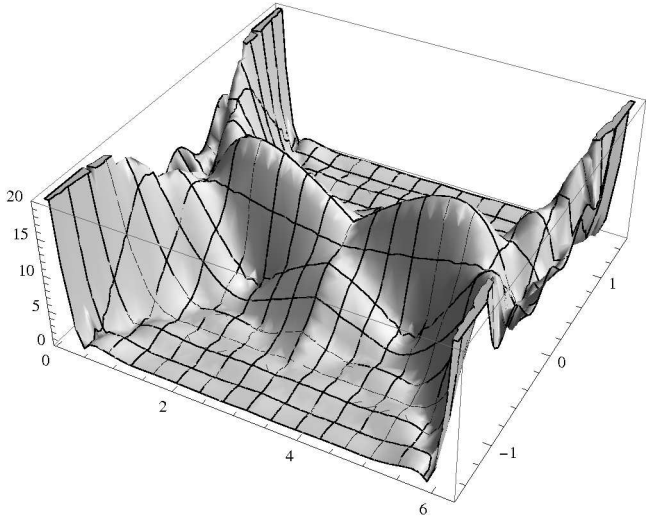}
\caption{Wave packet $|\psi(b,\phi)|$ formed out of the solutions (\ref{hypersol}), with $\lambda=0.01$ and $\nu_0=3\pi G =1$, so that $b\in[0,2\pi)$.}
\label{hypergplot}
\end{figure}

\begin{figure}[ht]

\includegraphics[scale=0.67]{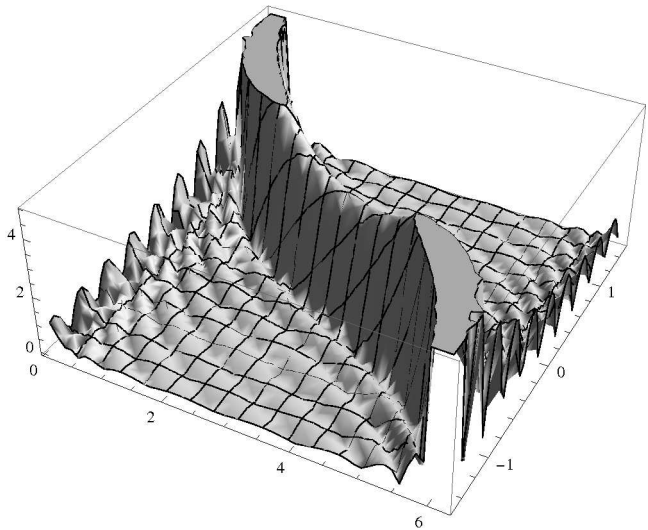}
\caption{Wave packet $|\psi(b,\phi)|$ in solvable LQC, with $A_{\rm L}(k)=0,\;A_{\rm R}(k)=e^{-0.01k^2}$ in (\ref{solution}), where again $\nu_0=3\pi G=1$.}
\label{slqcplot}
\end{figure}

For the operator choice (\ref{symme}), the free equation is
\bena
\partial_\phi^2\psi-6\pi G\left\{\left[\partial_b\frac{2}{\nu_0}\sin\left(\frac{b\,\nu_0}{2}\right)\right]^2\right.&&\nonumber
\\+\left. \left[\frac{2}{\nu_0}\sin\left(\frac{b\,\nu_0}{2}\right)\partial_b\right]^2\right\}\psi&=&0\,.
\eena
One can make the substitution
\ben\label{zb}
z=\sin\left(\frac{b \nu_0}{2}\right)
\een
to bring this to the form
\bena
\partial^2_\phi\psi-6\pi G\left[(1-2z^2)+(4z-6z^3)\partial_z\right.&&\nonumber
\\\left.+2z^2(1-z^2)\partial_z^2\right]\psi&=&0\,.\label{nokg}
\eena
By separation of variables and an \emph{Ansatz} $\psi(z,\phi)=e^{\im k \phi}z^\mu\,g(z^2)$, one obtains ($y:= z^2$)
\bena
\left\{\frac{1}{8y}\left[1+\frac{k^2}{6\pi G}+2(\mu+\mu^2)\right]-\frac{1}{4}(1+\mu)^2\right\}g(y)&&
\label{mueq}
\\+\left[\left(\mu+\frac{3}{2}\right)-y(\mu+2)\right]g'(y)+y(1-y)g''(y)&=&0\,,\nonumber
\eena
which the reader will recognize as a special case of Euler's hypergeometric differential equation \cite[Eq.~9.151]{GR}
\ben
y(1-y)g''+[c-(a+b+1)y]g'-ab\,g=0\,,
\label{hypergeom}
\een
provided that the coefficient of $g(y)/y$ is made to vanish:
\ben
\mu_{\pm}=-\half\pm\frac{\im}{2}\sqrt{1+\frac{k^2}{3\pi G}}\,.
\label{mus}
\een
Then, Eq.~(\ref{mueq}) is (\ref{hypergeom}) with $a=b=(\mu+1)/2$ and $c=\mu+3/2$; two independent solutions around $y=0$ are given in terms of Gaussian hypergeometric functions
\ben
\psi_{\mu_\pm}=e^{\im k \phi}z(b)^{\mu_\pm}\;{}_2 F_1\left[\frac{\mu_\pm+1}{2},\frac{\mu_\pm+1}{2};\mu_\pm+\frac{3}{2};z(b)^2\right],
\label{hypersol}
\een
where $z(b)$ is given by Eq.~\Eq{zb} and $\mu_{\pm}$ are the two roots (\ref{mus}). In terms of the variables $\nu$ and $\phi$, the general solution can then be written as
\bena
\psi(\nu,\phi)&=&\frac{\nu_0}{2\pi}\int\limits_0^{2\pi/\nu_0} db\,e^{-\im\nu b}\int\limits_{-\infty}^{+\infty}dk\,
\\&&\times\left[A_+(k)\,\psi_{\mu_+(k)}(b,\phi)+A_-(k)\,\psi_{\mu_-(k)}(b,\phi)\right]\nonumber
\eena
for some functions $A_+(k)$ and $A_-(k)$ (the interpretation of which as left- and right-moving modes is less obvious). Taking $A_+=e^{-\lambda k^2}$ and $A_-=0$ we obtain a wave packet, plotted in Fig.\ \ref{hypergplot} (using {\sc Mathematica}) as a function of $b$ and $\phi$. The wave packet is peaked on both ``expanding'' and ``contracting'' classical solutions $\phi=\pm(12\pi G)^{-1/2}\ln\tan(\nu_0 b/4)$. This can be compared with a wave packet composed out of right-moving modes $x'(b)e^{\im k[\phi+x(b)]}$ in solvable LQC, sharply peaked on the ``contracting'' branch (see Fig.\ \ref{slqcplot}).

The bounce picture is not as clear, analytically, as that in sLQC. Equation \Eq{nokg} is not a Klein--Gordon equation, derivatives in $\phi$ do not map directly into derivatives in $z$, and the volume operator $\hat\nu\propto \sqrt{1-z^2}\partial_z$ acts on a function whose $z$ and $\phi$ dependence are completely uncorrelated. Figure \ref{hypergplot} suggests that also here the wave packet follows the classical trajectories [Eq.\ (\ref{cosh})] which contain a bounce, but we must leave a detailed numerical investigation of the existence of a bounce to future work.

\

For Wheeler--DeWitt quantum cosmology, the free field equation is just the $(1+1)$-dimensional wave equation with general solution $\psi(a,\phi)=\psi_+[\phi-\sqrt{3/(4\pi G)}\ln a]+\psi_-[\phi+\sqrt{3/(4\pi G)}\ln a]$, decomposed into Fourier modes as
\ben
\psi(a,\phi)=\int\limits_{-\infty}^{+\infty}dk\,\left\{f_{\rm L}(k)\,e^{\im k[\phi-{\cal N}(a)]}+f_{\rm R}(k)\,e^{\im k[\phi+{\cal N}(a)]}\right\}\,.
\een

\subsection{Propagator}
\label{propfock}

Having fixed a self-adjoint operator $\hat{\cK}$ in the action (\ref{akshn}), one needs to invert it to obtain the propagator of the theory. In the generic case (\ref{wdw}), we can give a ``spin foam'' series expansion by using the results given in \cite{2point} (extending those of \cite{ach1,ach2}), where we determined the Feynman propagator corresponding to a constraint of the form $-(\Theta+\partial_\phi^2)$ [that is, for $B(\nu)=1$] to be
\bena
\im G_{{\rm F}}&=&\sum_{M=0}^{\infty}\sum_{{\nu_{M-1},\ldots,\nu_1}\atop{\nu_m\neq\nu_{m+1}}}\Theta_{\nu\,\nu_{M-1}}\ldots\Theta_{\nu_2\,\nu_1}\Theta_{\nu_1\,\nu'}
\\&&\times\prod_{k=1}^p\frac{1}{(n_k-1)!}\left(\frac{\partial}{\partial\Theta_{w_k w_k}}\right)^{n_k-1}\nonumber
\\&&\left[\sum_{m=1}^p\frac{e^{-\im\sqrt{\Theta_{w_m w_m}}(\phi-\phi')}}{2\sqrt{\Theta_{w_m w_m}}\prod_{{j=1}\atop{j\neq m}}^p(\Theta_{w_m w_m}-\Theta_{w_j w_j})}\right]\,.\nonumber
\eena
We refer to \cite{2point} for notation and details of the calculation. From this it follows that a Green's function (right inverse) for $\hat{\cK}$ with general $B(\nu)$ is 
\ben
G_{{\rm R}}(\nu,\phi;\nu',\phi'):=\frac{G_{{\rm F}}(\nu,\phi;\nu',\phi')}{B(\nu')}\,, 
\een
since
\bena
\hat{\cK}_{\nu,\phi} G_{{\rm R}}(\nu,\phi;\nu',\phi')& = &\frac{B(\nu)}{B(\nu')}(-\Theta_\nu-\partial_\phi^2)G_{{\rm F}}(\nu,\phi;\nu',\phi')\nonumber
\\&=&\delta_{\nu,\nu'}\delta(\phi-\phi')\,,
\label{greeneq}
\eena
where the subscript of $\hat \cK$ indicates that differential and difference operators act on the first argument only. Since $\hat{\cK}$ is self-adjoint, a left inverse is obtained by taking the adjoint of the right inverse, $G_{{\rm L}}(\nu,\phi;\nu',\phi')=\overline{G_{{\rm R}}(\nu',\phi';\nu,\phi)}$.

For the sLQC model, we can give a more explicit expression for the propagator, exploiting the relation to the Klein--Gordon equation; a Green's function for (\ref{slqc}) is given by
\bena
G^{\mathfrak{s}}_{{\rm R}}(b,\phi;b',\phi')&=&\left.\frac{dx}{db}\right|_{b=b'}\,\im\partial_b\left\{G_{{\rm KG}}[x(b),\phi;x(b'),\phi']\right\}\nonumber
\\&=&\im\partial_b\left\{\nu_0\frac{G_{{\rm KG}}[x(b),\phi;x(b'),\phi']}{4\sqrt{3\pi G}\sin(\frac{\nu_0 b'}{2})}\right\}\,,
\eena
where $G_{{\rm KG}}$ is the Feynman propagator for the Klein--Gordon equation. Explicitly \cite{propref},
\ben
G_{{\rm KG}}=-\frac{1}{4\pi}\ln\left\{\mu^2[(\phi-\phi')^2-(x-x')^2-\im\epsilon]\right\}\,,
\label{kgprop}
\een
where the usual $\im\epsilon$ prescription cancels the singularities on the ``light cone,'' so that
\ben
G^{\mathfrak{s}}_{{\rm R}}=\frac{(96\pi^2 G)^{-1}\,\im\nu_0^2\,[x(b)-x(b')]}{\sin\frac{\nu_0 b}{2}\sin\frac{\nu_0 b'}{2}\left\{(\phi-\phi')^2-[x(b)-x(b')]^2-\im\epsilon\right\}}\,.\label{prpx}
\een
$G^{\mathfrak{s}}_{{\rm R}}$ blows up as $b\rightarrow 0$ or $b'\rightarrow 0$ because the coordinate transformation from $b,b'$ to $x,x'$ becomes singular there, as we noted below (\ref{transf}). Hence, these are not physical singularities. To obtain a Green's function which can be extended to those values, one would have to solve Eq.~(\ref{greeneq}) with appropriate boundary conditions. In the above example, the $b$ representation elects $\nu$ as the momenta; swapping representation, the physical interpretation of the light-cone poles as classical trajectories would be unchanged.

Notice that the choice of the Feynman propagator for the particular Green function to use can be justified formally by the analogy with the free particle case, and thus by a definition of ``time-ordering'' and thus causality conditions with respect to the values of the scalar field used as internal time for the system. Another justification, possibly more satisfactory, for the same choice of analytic continuation in the complex plane is the fact that this choice makes not only the propagator itself but also the formal field-theory path integral well defined, at least as far as the free theory is concerned.

\

In comparison, we note that for the Hamiltonian constraint (\ref{wdwcon}) of standard Wheeler--DeWitt quantum cosmology one just considers the Klein--Gordon equation, and the propagator of the theory is $G_{{\rm KG}}[\sqrt{3/(4\pi G)}\,\ln a,\phi;\sqrt{3/(4\pi G)}\,\ln a',\phi']$.


\section{Taking interactions into account: mean-field approximation}\label{mfa}

We now want to extend our analysis to the field interactions, i.e., to the effects of topology change on the dynamics of a single universe (or, in the alternative interpretation we suggested, of the interaction of the various homogeneous patches of the universe on the dynamics of each of them). As in any nontrivial field theory, the exact solution of the dynamics is beyond question, and one has to resort to approximation methods and various truncations. One is of course the perturbative expansion of the field theory around the Fock vacuum and the study of the corresponding Feynman diagrams and amplitudes. This is, for example, the level at which the current understanding of full GFT's (and of the corresponding spin foam models) stands. Another type of technique, aiming at an approximate understanding of nonperturbative features of the theory and at the extraction of effective dynamics from the ``fundamental'' one, is mean field theory. The application of such technique in the full GFT framework has just started \cite{meanfield}, and a similar study of the toy model we defined here is what we focus on in the following.

\
  
As said, a first approximation to the dynamics of an interacting system is obtained if one assumes that, in an appropriate quantum state $|\xi\rangle$, the system fluctuates around a configuration with nonvanishing expectation value of the field operator $\hat\Psi$ and replaces $\hat\Psi=\langle\hat\Psi\rangle+\delta\hat\Psi$ in the quantum (operator) version of the field equation (\ref{fieldeq}). A similar expansion can already be done at the classical level by writing the field $\Psi$ appearing in the action (\ref{intact}) as $\Psi=\Psi_0+\delta\Psi$, that is, when studying the field theory dynamics around a different (nontrivial) vacuum $\Psi_0$. The resulting effective action from such an expansion [starting from Eq.~(\ref{intact})] is the following:

\bena
S_{{\rm eff}} &=& S[\Psi = \Psi_0 + \delta\Psi] - S[\Psi=\Psi_0] \nonumber \\ &=& \sum_{\nu}\int d\phi\;\left[\Psi_0(\nu,\phi)+\half\delta\Psi(\nu,\phi)\right]\hat{\cK}\delta\Psi(\nu,\phi) \nonumber \\ && +\sum_{j=2}^n\frac{\lambda_j}{j!}\sum_{\nu_1\ldots\nu_j}\int d\phi_1\ldots d\phi_j\; f_j(\nu_i,\phi_i) \nonumber \\ &&\times \sum_{m=1}^j {j \choose m} \prod_{k=1}^{j-m}\Psi_0(\nu_k,\phi_k)\prod_{l>j-m}^j\delta\Psi(\nu_l,\phi_l)\,,\nonumber\\ \label{effect}
 \eena

where we have assumed that the functions $f_j(\nu_1,\ldots,\nu_j,\phi_i,\ldots,\phi_j)$ are symmetric under any permutations of their $j$ pairs of variables. This assumption, needed only to have a more compact expression for the result, can be lifted straightforwardly.\footnote{When the assumption is not satisfied, the expression replacing the third and fourth lines in (\ref{effect}), for given $j$, is obtained by: (i) choosing $m$ ordered elements out of the ordered set of $j$ variables of the functions $f_j$; (ii) convoluting $m$ fields $\delta\Psi$ and $(j-m)$ fields $\Psi_0$ with the same functions $f_j$, with respect to the chosen variables; (iii) summing over $m$ from $1$ to $j$.}

We can then isolate the terms that are linear, quadratic and higher-than-quadratic in the dynamical field $\delta\Psi$ to obtain (using the standard convention that a sum running from $n$ to $n-1$ is empty)

\bena
S_{{\rm eff}} &=& \left[ \sum_{\nu}\int d\phi\;\Psi_0(\nu,\phi)\hat{\cK}\delta\Psi(\nu,\phi)\,+ \right. \nonumber \\ &&\left.+ \sum_{j=2}^n\frac{\lambda_j}{(j-1)!}\sum_{\nu_1\ldots\nu_j}\int d\phi_1\ldots d\phi_j\right. \nonumber \\ 
&& \times\left.f_j(\nu_i,\phi_i)\left(\prod_{k=1}^{j-1}\Psi_0(\nu_k,\phi_k)\right)\delta\Psi(\nu_j,\phi_j)\right] \nonumber\\ 
&&+\half\left[ \sum_{\nu}\int d\phi\;\delta\Psi(\nu,\phi)\hat{\cK}\delta\Psi(\nu,\phi)\, \right. \nonumber\\ 
&&\left.+\, \sum_{j=2}^n\frac{\lambda_j}{(j-2)!}\sum_{\nu_1\ldots\nu_j}\int d\phi_1\ldots d\phi_j\,f_j(\nu_i,\phi_i)\right.\nonumber\\ 
&& \times\left.\,\left(\prod_{k=3}^{j}\Psi_0(\nu_k,\phi_k)\right)\delta\Psi(\nu_{1},\phi_{1})\delta\Psi(\nu_2,\phi_2)\right] \nonumber\\ 
&& +\left[\sum_{j=2}^n\frac{\lambda_j}{j!}\sum_{\nu_1\ldots\nu_j}\int d\phi_1\ldots d\phi_j\; f_j(\nu_i,\phi_i)\right.\nonumber\\
&&\times\left.\sum_{m=3}^j {j \choose m} \prod_{k=1}^{j-m}\Psi_0(\nu_k,\phi_k)\prod_{l>j-m}^j\delta\Psi(\nu_l,\phi_l)\right]\,.\nonumber\\ \label{effect2}
\eena
The term that is linear in $\delta\Psi$ vanishes if the mean-field configuration $\Psi_0$ is chosen to satisfy the classical equation of motion of the original field theory. In practice, it is extremely hard to find a classical solution of the interacting theory, so, as is the case in some condensed matter systems, we resort to a further approximation valid in the limit of small coupling constants. That is, we will choose solutions of the {\it free} field theory as our mean-field vacua $\Psi_0$, either exact (when possible) or approximate, and then assume that the coupling constants $\lambda_j$ are very small (that is, that topology change is strongly suppressed in this cosmological second-quantized toy model) and that, because of this, the same vacua represent approximate solutions of the full equation of motion. Under these assumptions, we can neglect the linear terms in the above effective action.

\

The quadratic term in $\delta\Psi$ defines an effective Hamiltonian constraint for the cosmological second-quantized model, taking into account the small processes of merging/splitting of homogeneous isotropic universes/patches. This effective Hamiltonian constraint operator is given by:
\ben
\hat{\cK}_{{\rm eff}}^{\Psi_0,f_j,\lambda_j}\,=\,\delta(\phi_1-\phi_2)\delta_{\nu_1,\nu_2}\hat{\cK}\,+\,\sum_{j=2}^n\frac{\lambda_j}{(j-2)!}\,\hat{\cK}^{\Psi_0,f_j,\lambda_j}_j
\een
with
\bena
\hat{\cK}^{\Psi_0,f_j,\lambda_j}_j&=& \sum_{\nu_3\ldots\nu_j}\int d\phi_3\ldots d\phi_j\, \prod_{k=3}^j\Psi_0(\nu_k,\phi_k)\nonumber \\ &&\times f_j(\nu_1,\nu_2,\ldots,\nu_j,\phi_1,\phi_2,\ldots,\phi_j)\,.
\eena
It depends on the original coupling constants $\lambda_j$, on the original interaction kernels $f_j$, and, crucially, on the mean-field configuration $\Psi_0$ chosen as new vacuum.

\

The new effective interactions $V_{{\rm eff}}$ for $\delta\Psi_0$ depend on the same data, and are given by the last term in Eq.~(\ref{effect2}):
\bena
V_{{\rm eff}}&=& \sum_{j=3}^n\frac{\lambda_j}{j!}\sum_{\nu_1\ldots\nu_j}\int d\phi_1\ldots d\phi_j\,f_j(\nu_i,\phi_i)  \\ && \times\sum_{m=1}^j {j \choose m} \prod_{k=1}^{j-m}\Psi_0(\nu_k,\phi_k)\prod_{l>j-m}^j\delta\Psi(\nu_l,\phi_l)\,, \nonumber
\eena
from which one can read out the new interaction kernels.

\

The above is totally general. The simplest case is when the original field theory contains only interactions of the lowest (nonquadratic) order, that is, when $n=3$. Then, one obtains the effective action (now assuming the linear term vanishes)
\bena
S_{{\rm eff}}[\delta\Psi] &=& \half\sum_{\nu_1,\nu_2}\int d\phi_1\,d\phi_2\,\delta\Psi(\nu_1,\phi_1)\,\hat{\cK}_{{\rm eff}}\,\delta\Psi(\nu_2,\phi_2)\nonumber\\ 
&&+\frac{\lambda_3}{3!} \sum_{\nu_1,\nu_2,\nu_3}\!\int d\phi_1\,d\phi_2\,d\phi_3\,f_3(\nu_i,\phi_i)\nonumber\\ 
&&\times\prod_{k=1}^3\delta\Psi(\nu_k,\phi_k)\,,
\eena
with the effective Hamiltonian constraint operator
\bena
\hat{\cK}_{{\rm eff}}\,&=&\,\delta(\phi_1-\phi_2)\delta_{\nu_1,\nu_2}\hat{\cK} +\lambda_2\,f_2(\nu_1,\phi_1;\nu_2,\phi_2) \nonumber \\ && +\lambda_3\sum_{\nu_3}\int d\phi_3 \,f_3(\nu_i,\phi_i)\,\Psi_0(\nu_3,\phi_3)\,.\label{heff}
\eena
In this simple case, one can also easily give the generalization to nonsymmetric interaction kernels: 
\bena
\hat{\cK}_{{\rm eff}}\,&=&\,\hat{\cK}(\nu_1,\phi_1;\nu_2,\phi_2) +\half\left\{\vphantom{\half}\lambda_2\,f_2(\nu_1,\nu_2,\phi_1,\phi_2)\,\right.\nonumber \\ &&+\frac{\lambda_3}{3}\sum_{\nu_3}\int d\phi_3 \,\left[f_3(\nu_1,\nu_2,\nu_3,\phi_1,\phi_2,\phi_3)\right. \nonumber \\ &&+ f_3(\nu_1,\nu_3,\nu_2,\phi_1,\phi_3,\phi_2)\nonumber \\ 
&&+ \left.f_3(\nu_3,\nu_2,\nu_1,\phi_3,\phi_2\phi_1)\right]\Psi_0(\nu_3,\phi_3)\nonumber\\
&&+\left.\vphantom{\half}(\nu_1\leftrightarrow \nu_2)\right\}.
\eena

\

It is clear that the main issue in this approach, for the extraction of effective single-universe dynamics from the initial field-quantized model, is the choice of mean field $\Psi_0$. We have already anticipated the general requirement of choosing (approximate) solutions of the original field theory equations, and the difficulty involved in doing so. We will now work out a choice of mean-field vacuum. Having done so, we will turn to the role of the original interaction kernels $f_j$ and the consequences, at the level of the effective Hamiltonian constraint, of some interesting choices of the same. We keep this final discussion on the possible resulting dynamics rather brief, leaving a more thorough analysis to the future.\footnote{A remark is in order. The logic of a mean-field approximation is the following. One starts from a given dynamical model of the universe in second quantization through some functions $f_j$ (in addition to a given free theory dynamics). One aims at obtaining an effective free theory, taking into account some effects of the presence of interactions, around a new vacuum. One has then to identify what this relevant new vacuum is, and extract the effective dynamics around it (which will of course depend on the original choice of interactions $f_j$). In our presentation we will be forced to follow a different logic. We will first discuss choices of mean field, then consider interesting possibilities for effective Hamiltonian dynamics one may want to obtain as a result of the mean-field approximation of the second-quantized dynamics, and finally we will discuss which initial model (functions $f_j$) would lead to the effective Hamiltonian considered, given a mean-field configuration. The reason for this line of argument is twofold. First, given the progress in the context of LQC and the novelty of our second-quantized reformulation, we have some control over the possible forms of mean-field configurations but very little constraints on the ``correct'' choice of interactions. Second, in the present paper, whose main goal is to {\it introduce} the new second-quantized framework, we are more interested in exploring the available possibilities, the outcomes of various choices of models, and the ways to deal with them, rather than analyzing the properties of one specific model.}

\

One can follow two conceptually different but mathematically similar approaches to the choice of mean-field vacuum, yielding, eventually, the same result. The first would be to take a known physical state from the first-quantized theory and write
\ben\label{vev}
\Psi_0(\nu,\phi)=\psi(\nu,\phi)\,.
\een
In the second, one takes a second-quantized coherent state in a Fock space picture \cite{altland},
\ben
|\xi\rangle=\exp\left[\int\,d\varrho(k)\,\xi(k)\,a^{\dagger}_k\right]|0\rangle\,,
\een
where $k$ labels classical solutions to the constraint, $d\varrho(k)$ is some measure determined by the normalization of single-particle states $|k\rangle = a^{\dagger}_k|0\rangle$ [by imposing $\int\,d\varrho(k)\langle k|k'\rangle=1$], and $|0\rangle$ is the Fock vacuum. These states are eigenstates of the annihilation operators $a_k$ with eigenvalue $\xi(k)$, so that the field operator $\hat\Psi(\nu,\phi)$ has expectation value
\ben
\langle\xi|\hat\Psi(\nu,\phi)|\xi\rangle = \left[\int\,d\varrho(k)\,\xi(k)\chi_k(\nu,\phi)+\cc\right]\|\xi\|^2\,,
\een
where $\chi_k(\nu,\phi)$ is the solution to Eq.~(\ref{wdw}) labeled by $k$. Clearly, this expectation value can then be equivalently viewed as a first-quantized (real) wavefunction $\psi(\nu,\phi)$, Eq.~\Eq{vev}. In both viewpoints there is a normalization condition: For $|\xi\rangle$ to be in the Fock space, one must have $\int d\varrho(k)|\xi(k)|^2<\infty$, while $\psi(\nu,\phi)$ defines a first-quantized state if it has finite norm in the appropriate physical inner product, e.g.,
\ben
\|\psi\|^2=\sum_\nu B(\nu)|\psi(\nu,\phi_0)|^2\,,
\een
at some fixed $\phi_0$ \cite{improv}.

For the solvable sLQC model, where the general solution is Eq.~(\ref{solution}), we have to choose appropriate functions $A_{\rm L}(k)$ and $A_{\rm R}(k)$ in the mean-field approximation. If we consider only a single right-moving mode $k_0>0$ [$A_{\rm L}(k)=0$, $A_{\rm R}(k)=\delta(k-k_0)$],
\bena
\psi_{k_0}(\nu,\phi)&=&\frac{\nu_0}{2\pi}\int\limits_{-\infty}^{+\infty} dx\,e^{-\im\nu b(x)}e^{\im k_0(\phi+x)}+\cc\nonumber\\
&\sim&\cos\left[\frac{k_0}{\sqrt{12 \pi G}}\ln\left(\frac{2\sqrt{12 \pi G}\nu}{k_0\nu_0}\right)-\frac{\pi}{4}\right]\nonumber
\\&&\times\frac{2\nu_0}{(3\pi^3 Gk_0^2)^{1/4}}\cos(k_0\phi)\,.
\eena
Here we have used a stationary-phase approximation
\ben
\int dx\,e^{\im g(x)}\sim\sum_{g'(x_0)=0}e^{\im g(x_0)}\sqrt{\frac{2\pi}{\im g''(x_0)}}\nonumber
\een
in the limit $\nu\gg\nu_0$, in which we find that $\psi$ does not decay and, in general, does not satisfy a normalization condition. Alternatively, we might consider a Gaussian, giving a wave packet centered around a classical trajectory $\phi=-x=-(12\pi G)^{-1/2}\ln\tan(\nu_0 b/4)$, similar to the wave packets studied by Kiefer for Wheeler--DeWitt quantum cosmology \cite{kiefer}. The stationary-phase method then gives
\bena
\psi(\nu,\phi)&=&\frac{\nu_0}{2\pi}\int\limits_{-\infty}^{+\infty} dx\,dk\,e^{-\im\nu b(x)-\lambda(k-k_0)^2+\im k(\phi+x)}+\cc\nonumber
\\&=&\frac{\nu_0}{2\sqrt{\pi\lambda}}\int\limits_{-\infty}^{+\infty} dx\,e^{\im k_0(\phi+x)-\frac{1}{4\lambda}(\phi+x)^2-\im\nu b(x)}+\cc\nonumber
\\&\sim&\left(\frac{2\nu\sqrt{12 \pi G}}{k_0\nu_0}\right)^{-\frac{1}{48\pi G\lambda}\ln\left(\frac{2\nu\sqrt{12 \pi G}}{k_0\nu_0}\right)}\psi_{k_0}(\nu,\phi)\,,\nonumber\\
\eena
where we pick up an extra factor from the Gaussian. For large $\nu$, the field now falls off faster than any power of $\nu$. The scalar field $\phi$ not only acquires an effective periodic potential [$V\sim f(a)\cos(k_0\phi)$ for $n=3$], but it also becomes nonminimally coupled with gravity via a nontrivial function $f(a)$.

\

We can redo the analysis for Wheeler--DeWitt cosmology using a wave packet of the usual form,
\bena
\psi(a,\phi)  & = & \int\limits_{-\infty}^{+\infty}dk\,e^{-\lambda(k-k_0)^2}\,e^{\im k[\phi-{\cal N}(a)]}\nonumber
\\& = &\sqrt{\frac{\pi}{\lambda}}\, e^{\im k_0[\phi-{\cal N}(a)]}\,e^{-\frac{1}{4\lambda}[\phi-{\cal N}(a)]^2},
\eena
so that for large $a$ (and at fixed $\phi$) we find again a fall-off behavior
\ben
\psi(a,\phi) \sim a^{\sqrt{\frac{3}{4\pi G}}\frac{\phi}{2\lambda}-\frac{3}{16\pi G\lambda}\ln a}
\een
faster than any power of $a$.

\

In general, any statement about the explicit form of the contribution to $\hat{\cK}$ will strongly depend not only on the chosen field configuration $\Psi_0$, but also on the form of the interactions in our model, which are not strongly constrained. Let us discuss possible results for this effective Hamiltonian dynamics, and how to obtain them. 

A possibility which has been suggested by studies in LQC \cite{ach1,ach2} is that, for a ``monomial'' interaction, the GFT coupling constant $\lambda$ is related to the cosmological constant $\Lambda$. This possibility had been also considered previously \cite{ori1, ori2}, but, in analogy with matrix models and tensor models and with the ``third-quantization'' model of \cite{giddstrom}, the coupling $\lambda$ would be expected to be related to the {\it exponential of the cosmological constant} rather than $\Lambda$ itself. For example, the formal arguments of \cite{giddstrom} suggest that a second-quantized field-theory Feynman amplitude should correspond to $e^{\im S}$ evaluated on the classical spacetime represented by the Feynman diagram. 

Our simplified scheme suggests another way to obtain a relation between a fundamental field-quantized coupling and an effective cosmological constant. Under the assumption that our toy cosmological model comes out from some more fundamental GFT dynamics, and thus that the coupling constant of the toy model can be related to the fundamental GFT interactions, the mean field approximation can relate very directly the coupling constant of the cosmological model to an effective cosmological constant in the single-universe dynamics (free theory), that is, in an effective Hamiltonian constraint. Let us see how this can happen. 

The effective contribution to $\hat{\cK}$ should be of the form $\cK_\Lambda =\Lambda\,B(\nu)\,\nu^2$. This term grows with large $\nu$ and, for a trivial polynomial interaction (that is, trivial interaction kernels $f_j$), presumably it could not come from a normalizable $\Psi$. However, a general choice of interaction functional, such as that in Eq.~\Eq{intact}, can accommodate an effective cosmological constant. Actually, it can even reproduce a nonconstant scalar-field potential term,
\ben
\cK_V =V(\phi)\,B(\nu)\,\nu^2\,,
\een
for example of the type that would be needed for inflation in the early universe.

Take, again, $n=3$ and Eq.~\Eq{heff}. The contribution in $f_2$ is nontrivial only if the effective Hamiltonian is nonlocal, otherwise it would just be an extra piece defining the initial Hamiltonian constraint. Since we are interested in a local constraint, we can ignore it, set $\lambda_2=0$, and absorb $\lambda_3$ in the potential $V(\phi)$. Then, a function $f_3$ that happens to match the behavior of the mean-field vacuum and contains the appropriate dependence on $\nu$ and $\phi$ on top of it would be
\bena
f_3(\nu_i,\phi_i)&=& \delta(\phi_1-\phi_2)\delta(\phi_2-\phi_3)\delta_{\nu_1,\nu_2}\delta_{\nu_2,\nu_3}\nonumber\\
&&\times V(\phi_3)\,B(\nu_3)\nu_3^2[\Psi_0(\nu_3,\phi_3)]^{-1}\,.
\label{f3}
\eena
Notice that, by construction, $f_3$ is symmetric with respect to all its arguments.

Clearly, this is rather \emph{ad hoc} and one should have an independent justification (and possibly a full derivation from a fundamental GFT) for a given choice of interactions $f_3$, {\it such that} the wished-for effective Hamiltonian constraint comes out, {\it for a reasonable choice of nonperturbative vacuum} which should also be independently justified. 

However, the above derivation proves an intriguing possibility, to be explored further:  an underlying, more fundamental dynamics  of creation/annihilation of universes, i.e., topology change, or of merging/splitting of homogeneous and isotropic patches within a single universe could result, at an effective level, in a nontrivial potential term for the (homogeneous and isotropic) scalar field, and a cosmological constant term. 

Notice also that, in the first-quantized LQC and Wheeler--DeWitt frameworks, the presence of a potential spoils the separation of positive-frequency and negative-frequency sectors, as recalled, e.g., in \cite{2point}. This is not an issue in our second-quantized model (being a field theory), and we are able to generate a scalar potential (nontrivial in the inflationary early universe) without formally changing the structure of the free theory.

With the same procedure, we can obtain also other types of effective contributions, for instance a nonvanishing curvature $\textsc{k}=\pm 1$ (closed and open universe, respectively). It is sufficient to replace
\ben
\nu V(\phi)\to \nu V(\phi)-\nu^{1/3}\frac{3\textsc{k}}{8\pi G}
\een
in Eq.~\Eq{f3} to obtain an effective spatial curvature term of the form one finds in the classical Friedmann equation. It is also straightforward to obtain an effective Hamiltonian constraint corresponding to, e.g., the $\textsc{k}=1$ model in LQC \cite{kequalsone}. In fact also in that case the term corresponding to spatial curvature only acts by multiplication with a function of $\nu$.


\section{Discussion and outlook}

In this paper we outlined the construction of a field theory of universes drawing inspiration from the perspective advanced in quantum gravity by group field theory, and by the general idea of ``third quantization of gravity,'' that had been advanced in the early days of the subject \cite{giddstrom,kuchar,isham}. We have also studied various aspects of the formalism, in particular the consequences it has for the single-universe dynamics, that is, for the standard (loop) quantum cosmology setting. These come already from the embedding of the canonical dynamics within a field theory, as encoded in the free field theory. More interesting consequences, of course, come from the existence of interactions, which, we showed, can be taken into account via mean field approximation. Given the subject, and the current level of understanding of the fundamental theory (in either the LQG or in the GFT formulation), our goal was then in many respects necessarily of an exploratory nature. In particular, the least developed point of the discussion concerns the choice of interaction. As in all simplified models for cosmology, however, the main task will be to better justify the assumptions and the dynamics chosen from the fundamental theory, and to possibly show how such a simplified model can emerge naturally in some sectors of the full theory. Understanding this issue will also clarify the role and physical significance of conserved currents within the model.  

Before concluding, we should mention an alternative view of the physics of this ``group field cosmology,'' that we anticipated in passing in the course of our presentation. While loop quantum or Wheeler--DeWitt cosmology describe a single evolving quantum universe, in the field cosmological model one has many-particle interacting states. Instead of interpreting them as $n$ distinct universes merging and splitting in topology-changing ``scattering'' processes, one could think of them as $n$ FRW patches which, collected together, approximate a single inhomogeneous universe. This is reminiscent of the separate universe approach \cite{sepuniv,RS03a,LiL}, where inflationary large-wavelength perturbations are represented as spatial gradients among homogeneous patches of Hubble size.\footnote{For other work in the quantum cosmology context which is similar to the spirit of the separate universe approach, see, e.g., \cite{ed}} In each patch centered at some spatial point ${\bf x}$, one has a ``local'' scale factor $a(t,{\bf x})$, Hubble parameter $H(t,{\bf x})$, and so on. In particular, the local scale factor $a(t,{\bf x})=a(t)\,\exp[-\Phi_{\rm NL}(t,{\bf x})]$ encodes both the minisuperspace variable $a$ and the nonlinear scalar perturbation $\Phi_{\rm NL}$. Linear perturbations can then be identified with gradients. At the linear level, $a(t,{\bf x})\approx a(t)[1-\Phi_{\rm NL}(t,{\bf x})]$; call $\delta a= -a(t)\Phi_{\rm NL}(t,{\bf x})$. One has $[a(t,{\bf x}_1)-a(t,{\bf x}_2)]/(x_1^i-x_2^i)\sim \partial_i a(t,{\bf x})$, so that, up to a numerical factor, for a perturbation of wavelength $\lambda$ we get $\delta a\sim \lambda\partial_i a(t,{\bf x})$.

In our field-theory picture, the present-day universe would resemble some configuration of many particles (regions of linear size $b^{-1}$) that looks homogeneous to high precision at large scales. This could be a condensate phase of the theory where discrete translation invariance in the $\nu$ variable is spontaneously broken, as is known for systems in condensed matter physics. Then, the challenge would be to define a model whose collective behavior agrees with the standard cosmological perturbation theory. 

However, the resulting action would be nonlocal in order to accommodate the infinite multiplicity of spatial points into a finite-dimensional, minisuperspace-like phase space. For instance, following the above-mentioned gradient expansion, a linear perturbation would be defined via the ``interaction'' of two patches in a nonlocal quadratic term:
\ben\nonumber
\int da\, d\phi\int da'\, d\phi'\, f(a,a',\phi,\phi')\Psi(a,\phi)\Psi(a',\phi')\,,
\een
where $f$ is a function which should encode the correct dynamics to match with the perturbed Hamiltonian constraint. Just as we did for the effective Hamiltonian constraint in the previous section, one could explicitly calculate $f$ from this perturbed Hamiltonian constraint. The main difficulty to overcome, in developing properly the separate universe perspective of our second-quantized cosmology, is to develop first a proper quantum cosmology version of the separate universe approach to cosmological perturbations. Admittedly, it is unclear at the present stage whether applying our framework to the problem of cosmological perturbations has practical advantages over conventional strategies, unless some conservation law be implemented. It is yet another possibility worth exploring, however. On the other hand, the implementation of the separate universe idea within a field-theoretic formalism like the one we propose could have a better chance of being derived from fundamental formulations of quantum gravity such as GFT. This would give a more solid ground to this way of dealing with cosmological perturbations and also offer a way to test fundamental models via their cosmological predictions.

\section*{Acknowledgments}

This research was supported by Perimeter Institute for Theoretical Physics. Research at
Perimeter Institute is supported by the Government of Canada through Industry Canada and
by the Province of Ontario through the Ministry of Research \& Innovation. Support from the A.\ von Humboldt Stiftung, via a Sofja Kovalevskaja Award, is gratefully acknowledged.



\begin{thebibliography}{99}
\bibitem{libro}   D.\ Oriti (ed.), {\it Approaches to Quantum Gravity} (Cambridge University Press, Cambridge, U.K., 2009).
\bibitem{rovelli} C. Rovelli, {\it Quantum Gravity} (Cambridge University Press, Cambridge, U.K., 2007).
\bibitem{thi01} T.~Thiemann, \emph{Modern Canonical Quantum General Relativity} (Cambridge University Press, Cambridge, U.K., 2007); T.~Thiemann, \oarX{gr-qc/0110034}.
\bibitem{giddstrom} S.B.\ Giddings and A.\ Strominger, \doin{10.1016/0550-3213(89)90353-2}{Nucl.\ Phys.\ B {\bf 321},} \doin{10.1016/0550-3213(89)90353-2}{481
(1989)}.
\bibitem{3rdquant} S.\ Coleman, \doin{10.1016/0550-3213(88)90097-1}{Nucl.\ Phys.\ B {\bf 310}, 643 (1988)}; T. Banks, \doin{10.1016/0550-3213(88)90455-5}{Nucl.\ Phys.\ B {\bf 309}, 493 (1988)}; V.A.~Rubakov and P.G.~Tinyakov, \doin{10.1016/0370-2693(88)91373-1}{Phys.\ Lett.\ B {\bf 214}, 334 (1988)}; M.\ McGuigan, \doin{10.1103/PhysRevD.38.3031}{Phys.\ Rev.\ D {\bf 38}, 3031 (1988)}; A.~Hosoya and M.~Morikawa, \doin{10.1103/PhysRevD.39.1123}{Phys.\ Rev.\ D {\bf 39}, 1123 (1989)}; M.~McGuigan, \doin{10.1103/PhysRevD.39.2229}{Phys.\ Rev.\ D {\bf 39}, 2229 (1989)};  W.~Fischler, I.R.~Klebanov, J.~Polchinski, and L.~Susskind, \doin{10.1016/0550-3213(89)90290-3}{Nucl.\ Phys.\ B {\bf 327}, 157 (1989)}; Y.-M.~Xiang and L.~Liu,
\doin{10.1088/0256-307X/8/1/014}{Chin.\ Phys.\ Lett.\ {\bf 8}, 52 (1991)}; H.J.~Pohle, \doin{10.1016/0370-2693(91)90324-J}{Phys.\ Lett.\ B {\bf 261}, 257 (1991)}; S.~Abe, \doin{10.1103/PhysRevD.47.718}{Phys.\ Rev.} \doin{10.1103/PhysRevD.47.718}{D\ {\bf 47},\ 718\ (1993)}; T.~Horiguchi, \doin{10.1103/PhysRevD.48.5764}{Phys.\ Rev.\ D {\bf 48},} \doin{10.1103/PhysRevD.48.5764}{5764 (1993)};  M.A.~Castagnino, A.~Gangui, F.D.~Mazzitelli, and I.I.~Tkachev, \doin{10.1088/0264-9381/10/12/008}{Class.\ Quantum Grav.\ {\bf 10}, 2495} \doin{10.1088/0264-9381/10/12/008}{(1993)}; A.~Vilenkin,
\doin{10.1103/PhysRevD.50.2581}{Phys.\ Rev.\ D {\bf 50}, 2581 (1994)} [\oarX{gr-qc/9403010}]; L.O.~Pimentel and C.~Mora, \doin{10.1016/S0375-9601(01)00048-2}{Phys.} \doin{10.1016/S0375-9601(01)00048-2}{Lett.\ A {\bf 280}, 191 (2001)} [\oarX{gr-qc/0009026}]; S.~Gielen and D.~Oriti, in {\it Quantum Field Theory and Gravity}, edited by F.~Finster {\it et al.} (Springer, Basel, Switzerland, 2012) [\arX{1102.2226}].
\bibitem{FK}    E.~Fuchs and M.~Kroyter, 
  \doin{10.1016/j.physrep.2011.01.003}{Phys.\ Rep.\ {\bf 502}, 89 (2011)} [\arX{0807.4722}].
\bibitem{cuta5} G.\ Calcagni and G.\ Nardelli, 
  \doin{10.1016/j.nuclphysb.2009.08.004}{Nucl.\ Phys.\ B {\bf 823}, 234} \doin{10.1016/j.nuclphysb.2009.08.004}{(2009)} [\arX{0904.3744}].
\bibitem{cuta7} G.~Calcagni and G.~Nardelli,  
  \doin{10.1007/JHEP02(2010)093}{JHEP\ {\bf 1002}, 093  (2010)} [\arX{0910.2160}].
\bibitem{kuchar} K.\ Kuchar, in {\it Proceedings of the Fourth Canadian
Conference on General Relativity and Relativistic Astrophysics}, edited by
G.~Kunstatter {\it et al.} (World Scientific, Singapore, 1992); 
\doin{10.1142/S0218271811019347}{Int.\ J.\ Mod.\ Phys.\ Proc.\ Suppl.\ D {\bf
20},} \doin{10.1142/S0218271811019347}{3 (2011)}.
\bibitem{isham} C.\ Isham, in {\it Integrable systems, quantum groups, and quantum field theories}, 
edited by L.A.\ Ibort and M.A.\ Rodr\'{i}guez (Kluwer, Dordrecht, The Netherlands, 1993) [\oarX{gr-qc/9210011}].
\bibitem{Ori01} D.~Oriti,
  \doin{10.1088/0034-4885/64/12/203}{Rep.\ Prog.\ Phys.\ {\bf 64}, 1489 (2001)} [\href{http://arxiv.org/abs/gr-qc/0106091}{\ttfamily\com gr-qc/} \href{http://arxiv.org/abs/gr-qc/0106091}{\ttfamily\com 0106091}].
\bibitem{Per03}  A.~Perez, 
  \doin{10.1088/0264-9381/20/6/202}{Class.\ Quantum Grav.\ {\bf 20}, R43 (2003)} [\href{http://arxiv.org/abs/gr-qc/0301113}{\ttfamily\com gr-qc/} \href{http://arxiv.org/abs/gr-qc/0301113}{\ttfamily\com 0301113}].
\bibitem{ori1}  D.\ Oriti, 
in \cite{libro} 
[\oarX{gr-qc/0607032}].
\bibitem{ori2}   D.\ Oriti, in 
{\it Quantum Gravity}, edited by  B.\ Fauser {\it et al.} (Birkh\"auser, Basel, Switzerland, 2008) 
 [\oarX{gr-qc/0512103}].
\bibitem{ori3} D.\ Oriti, in {\it Foundations of Space and Time: Reflections on Quantum Gravity}, edited by J. Murugan {\it et al.} (Cambridge University Press, Cambridge, U.K., 2012) [\arX{1110.5606}].
\bibitem{tensor} R.\ Gurau and J.P.\ Ryan,
\doin{10.3842/SIGMA.2012.020}{SIGMA {\bf 8}, 020 (2012)} [\href{http://arxiv.org/abs/1109.4812}{\ttfamily\com arXiv:} \href{http://arxiv.org/abs/1109.4812}{\ttfamily\com 1109.4812}]; V.\
Rivasseau, [\arX{1112.5104}].
\bibitem{AsS}    A.\ Ashtekar and P.\ Singh,
\doin{10.1088/0264-9381/28/21/213001}{Class.\ Quantum Grav.\ {\bf
28},} \doin{10.1088/0264-9381/28/21/213001}{213001 (2011)} [\arX{1108.0893}].
\bibitem{BCM}    K.\ Banerjee, G.\ Calcagni, and M.\ Mart\'{\i}n-Benito,
\doin{10.3842/SIGMA.2012.016}{SIGMA} \doin{10.3842/SIGMA.2012.016}{{\bf 8}, 016 (2012)} [\arX{1109.6801}].
\bibitem{ach1}   A.~Ashtekar, M.~Campiglia, and A.~Henderson,  
  \doin{10.1016/j.physletb.2009.10.042}{Phys.} \doin{10.1016/j.physletb.2009.10.042}{Lett.\ B {\bf 681}, 347 (2009)} [\arX{0909.4221}].
\bibitem{ach2}   A.\ Ashtekar, M.\ Campiglia, and A.\ Henderson, 
 	\doin{10.1088/0264-9381/27/13/135020}{Class.} \doin{10.1088/0264-9381/27/13/135020}{Quantum Grav.\ {\bf 27}, 135020 (2010)} [\arX{1001.5147}].
\bibitem{2point} G.\ Calcagni, S.\ Gielen, and D.\ Oriti, \doin{10.1088/0264-9381/28/12/125014}{Class.\ Quantum} \doin{10.1088/0264-9381/28/12/125014}{Grav.\ {\bf 28}, 125014 (2011)} [\arX{1011.4290}].
\bibitem{improv} A.\ Ashtekar, T.\ Pawlowski, and P.\ Singh, 
  \doin{10.1103/PhysRevD.74.084003}{Phys.\ Rev.\ D} \doin{10.1103/PhysRevD.74.084003}{{\bf 74}, 084003 (2006)} [\arX{gr-qc/0607039}].
\bibitem{acs}    A.\ Ashtekar, A.\ Corichi, and P.\ Singh, \doin{10.1103/PhysRevD.77.024046}{Phys.\ Rev.\ D {\bf 77},} \doin{10.1103/PhysRevD.77.024046}{024046 (2008)} [\arX{0710.3565}].
\bibitem{dowker} F.\ Dowker, \oarX{gr-qc/0206020}.
\bibitem{selfad} W.\ Kami\'{n}ski and J.\ Lewandowski, \doin{10.1088/0264-9381/25/3/035001}{Class.\ Quantum} \doin{10.1088/0264-9381/25/3/035001}{Grav.\ {\bf 25}, 035001 (2008)} [\arX{0709.3120}].
\bibitem{hawpage} S.W.~Hawking and D.N.~Page, \doin{10.1016/0550-3213(86)90478-5}{Nucl.\ Phys.\ B {\bf 264}, 185} \doin{10.1016/0550-3213(86)90478-5}{(1986)}.
\bibitem{GFTdiffeos} A.\ Baratin, F.\ Girelli, and D.\ Oriti, \doin{10.1103/PhysRevD.83.104051}{Phys.\ Rev.\ D {\bf 83},} \doin{10.1103/PhysRevD.83.104051}{104051 (2011)} [\arX{1101.0590}].
\bibitem{GR}      I.S.~Gradshteyn and I.M.~Ryzhik, {\it Table of Integrals, Series, and Products} (Academic Press, London, U.K., 2007).
\bibitem{propref} N.D.\ Birrell and P.C.W.\ Davies, {\it Quantum Fields in Curved Space} (Cambridge University Press, Cambridge, U.K., 1982).
\bibitem{meanfield} D.\ Oriti and L.\ Sindoni, \doin{10.1088/1367-2630/13/2/025006}{New J.\ Phys.\ {\bf 13}, 025006 (2011)} [\arX{1010.5149}]; E.R.\ Livine, D.\ Oriti, and J.P.\ Ryan, \doin{10.1088/0264-9381/28/24/245010}{Class.\ Quantum Grav.\ {\bf 28}, 245010 (2011)} [\href{http://http://arxiv.org/abs/1104.5509}{\ttfamily\com arXiv:1104.} \href{http://http://arxiv.org/abs/1104.5509}{\ttfamily\com 5509}]; W.\ Fairbairn and E.R.\ Livine, \doin{10.1088/0264-9381/24/20/021}{Class.\ Quantum} \doin{10.1088/0264-9381/24/20/021}{Grav.\ \textbf{24}, 5277 (2007)} [\oarX{gr-qc/0702125}]; F.\ Girelli, E.R.\ Livine, and D.\ Oriti, \doin{10.1103/PhysRevD.81.024015}{Phys.\ Rev.\ D {\bf 81}, 024015 (2010)} [\arX{0903.3475}].
\bibitem{altland} A.\ Altland and B.\ Simons, {\it Condensed Matter Field Theory} (Cambridge University Press, Cambridge, U.K., 2006).
\bibitem{kiefer}  C.\ Kiefer, \doin{10.1103/PhysRevD.38.1761}{Phys.\ Rev.\ D {\bf 38}, 1761 (1988)}.
\bibitem{kequalsone} A.\ Ashtekar, T.\ Pawlowski, P.\ Singh, and K.\ Vandersloot, \doin{10.1103/PhysRevD.75.024035}{Phys.\ Rev.\ D {\bf 75}, 024035 (2007)} [\oarX{gr-qc/0612104}].
\bibitem{sepuniv} D.\ Wands, K.A.\ Malik, D.H.\ Lyth, and A.R.\ Liddle, \doin{10.1103/PhysRevD.62.043527}{Phys.\ Rev.\ D {\bf 62}, 043527 (2000)} [\oarX{astro-ph/0003278}].
\bibitem{RS03a} G.I.~Rigopoulos and E.P.S.~Shellard,
  \doin{10.1103/PhysRevD.68.123518}{Phys.\ Rev.\ D {\bf 68},} \doin{10.1103/PhysRevD.68.123518}{123518 (2003)} [\oarX{astro-ph/0306620}].
\bibitem{LiL}     D.H.\ Lyth and A.R.\ Liddle, \textit{The Primordial Density Perturbation} (Cambridge University Press, Cambridge, U.K., 2009).
\bibitem{ed} E.\ Wilson-Ewing, \arX{1108.6265}.
\end{thebibliography}
\end{document}